\documentclass[sigconf]{acmart}
\renewcommand\footnotetextcopyrightpermission[1]{} 
\usepackage[
    type={CC},
    modifier={by},
    version={4.0},
]{doclicense}
\usepackage{algorithm}
\usepackage{algpseudocode}

\AtBeginDocument{%
  }

\copyrightyear{2024}

\begin{document}
\title{TEGRA - Scaling Up Terascale Graph Processing with Disaggregated Computing}

\author{William Shaddix}
\author{Mahyar Samani}
\author{Marjan Fariborz}
\author{S.J. Ben Yoo}
\author{Jason Lowe-Power}
\author{Venkatesh Akella}
\affiliation{%
  \institution{University of California, Davis}
  \streetaddress{1 Shields Ave.}
  \city{Davis}
  \state{California}
  \country{USA}
  \postcode{95616}
}

\renewcommand{\shortauthors}{}



\begin{abstract}
Graphs are essential for representing relationships in various domains, driving modern AI applications such as graph analytics and neural networks across science, engineering, cybersecurity, transportation, and economics.
However, the size of modern graphs are rapidly expanding, posing challenges for traditional CPUs and GPUs in meeting real-time processing demands.
As a result, hardware accelerators for graph processing have been proposed.
However, the largest graphs that can be handled by these systems is still modest often targeting Twitter graph(1.4B edges approximately).
This paper aims to address this limitation by developing a graph accelerator capable of terascale graph processing.
Scale out architectures, architectures where nodes are replicated to expand to larger datasets, are natural for handling larger graphs.
We argue that this approach is not appropriate for very large scale graphs because it leads to under utilization of both memory resources and compute resources.
Additionally, vertex and edge processing have different access patterns.
Communication overheads also pose further challenges in designing scalable architectures.
To overcome these issues, this paper proposes TEGRA, a scale-up architecture for terascale graph processing.
TEGRA leverages a composable computing system with disaggregated resources and a communication architecture inspired by Active Messages.
By employing direct communication between cores and optimizing memory interconnect utilization, TEGRA effectively reduces communication overhead and improves resource utilization, therefore enabling efficient processing of terascale graphs.
\end{abstract}


\maketitle
\section{Introduction}
Graphs capture relationships between entities naturally and form the backbone of modern AI applications in the form of graph analytics and graph neural networks. Graphs are used in many applications within science, engineering, cybersecurity, transportation, and economics.
The size of the graphs keeps growing as the amount of data collected grows.
CPUs and GPUs cannot keep up with the requirements (especially real-time processing) in many applications.
As a result, hardware accelerators for graph processing has become a thriving research topic. 

Numerous hardware accelerators for graph processing such as PolyGraph~\cite{PolyGraph}, GraphPulse~\cite{GraphPulse}, Dalorex~\cite{Dalorex}, Graphicianado~\cite{Graphicionado}, ScalaBFS~\cite{scalabfs}, and ScalaGraph~\cite{scalagraph} have been proposed recently.
These systems can only handle modestly sized graphs, most of which targeting the Twitter graph (1.4B edges approximately). 
The goal of this work is to develop graph accelerators that can handle graphs that are orders of magnitude larger.
We are interested in terascale graph processing, i.e., the ability to handle trillions of edges.
Scale out architectures are natural for handling larger graphs.
This involves replicating each accelerator node and connecting the nodes with a high-bandwidth interconnection network.

We argue that this approach is not appropriate for very large scale graphs because it leads to the {\em stranding} of memory resources~(both capacity and bandwidth)  and  compute resources because 
graph processing requires both very high memory bandwidth (poor locality and poor data reuse) and very high capacity (to store trillions of edges).
So a simple scale-out architecture results in serious under utilization of the provisioned resources.
In addition, vertex processing requires random memory accesses while edge processing has some streaming access patterns.
This has to be matched with the characteristics and constraints of the underlying  memory (DRAM, HBM) technology and the memory access patterns during vertex and edge accessing.
DDRx  provides higher capacity  and lower bandwidth.
HBM provides lower capacity but higher bandwidth and reasonable for streaming access.
So, scaling out with either technology either leaves “stranded” bandwidth or “stranded” capacity.
Moreover, as the computation per vertex or edge accesses varies with different algorithms, there is a possibility of stranding computing capacity.
Furthermore, scaling out also couples the compute and memory resources making it difficult to address the stranding of memory or compute. 
 
 The coupling and under utilization of resources leads us to the conclusion that terascale graph processing calls for a composable computing systems with disaggregated resources in order to avoid the stranding of resources. 
We will describe our initial design of such a graph processing architecture called TEGRA - that is a scale-up architecture for terascale scale graph processing.
Specifically we show how we can avoid memory stranding with a disaggregated memory and interconnection stranding by taking advantage of communications architecture that is inspired by Active Messages~\cite{ActiveMessages}.
This helps TEGRA in two ways. It reduces the overhead of handling small messages through direct communication between the nodes as opposed to communicating through memory, and improves the utilization of the memory interconnect. We are able to improve the utilization of the memory interconnect because we have essentially two different interconnect networks, one optimized for memory accesses and one for the  messages exchanged between the vertices.
In this paper we describe our preliminary investigation into the design and evaluation of TEGRA.
The contributions of this paper are:
\begin{itemize}
    \item Design of a scale-out architecture for large scale graph processing.
    \item Evaluation of different memory technologies and memory architectures to help us with design space exploration of scalable graph processing architectures based on disaggregated resources.
\end{itemize}
\begin{figure}
    \includegraphics[scale=.35]{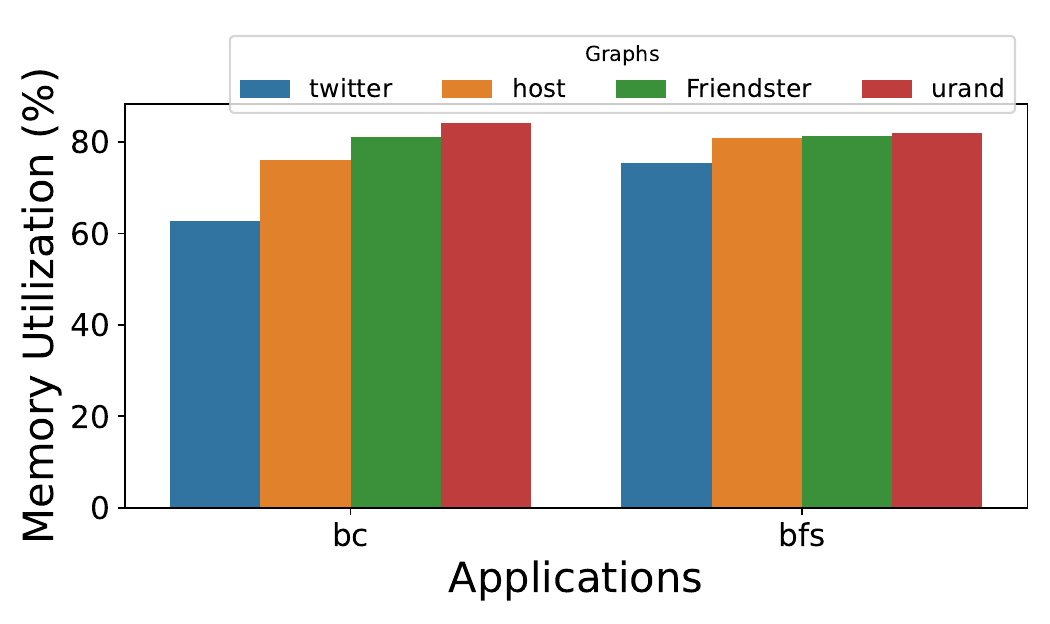}

    \caption{Bandwidth Utilization in different sized graphs on a scale out graph accelerator running the graph applications breadth first search~(bfs) and betweeness centrality~(bc)}
    \label{fig:diff-graphs}
\end{figure}

\begin{figure}
    \includegraphics[scale=.35]{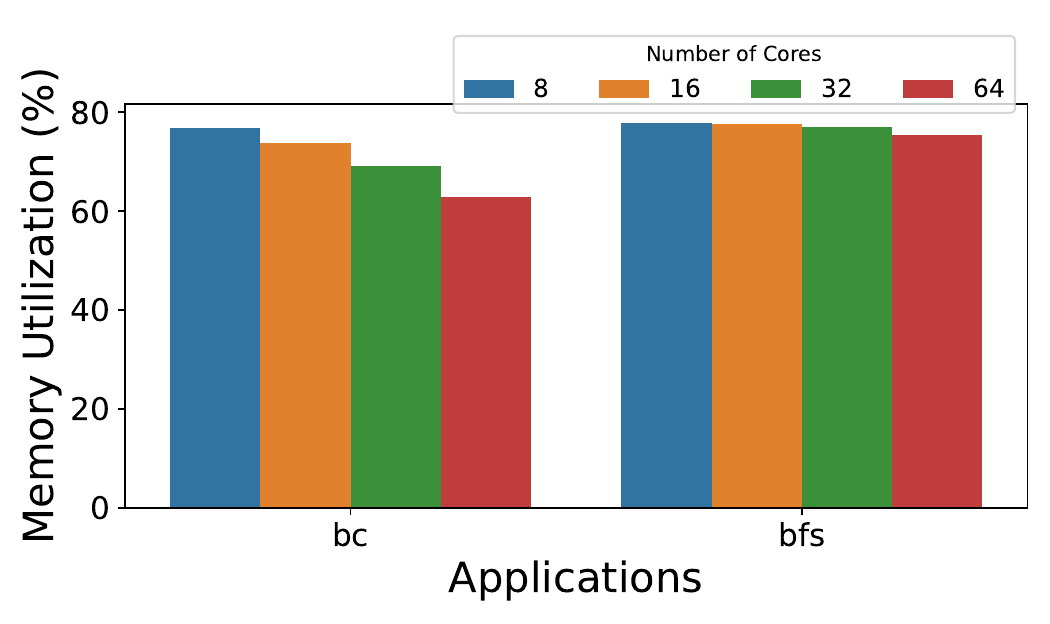}

    \caption{Bandwidth Utilization with different amounts of cores on a scale out graph accelerator}
    \label{fig:diff-cores}

\end{figure}

\section{Motivation}
The goals of this paper are to address the needs of modern scalable graph accelerators with regard to both functionality and performance of large graphs.
From a functional standpoint, hardwired accelerators are limited in which workloads and types of graphs they can process.
Many accelerators are unable to do more intense graph applications such as using dynamic graphs and triangle counting.
It has previously been shown that performance in balanced graph architectures is directly related to the memory bandwidth achieved~\cite{CAL}.
Figure ~\ref{fig:diff-graphs} shows a scale out graph accelerator's memory bandwidth utilization based on the size of the graph.
Even with large graphs, the accelerator only averages ~80\% bandwidth utilization,
meaning there is a theoretically untapped 20\% in performance from the memory system.
This bandwidth stranding comes from two potential sources, first the compute units in these systems are not able to send enough memory requests to fill the memory system. Second these compute units do not have work for intervals of the graph processing. 

Addressing the first cause we need  to increase compute throughput to send out more memory requests. Given that graphs are simple arithmetically, increasing the number of cores  improves compute throughput.
However, due to the coupling of memory and compute, increasing cores in a scale out accelerator also means increasing the amount of memory, leading to more memory bandwidth stranding.
Figure~\ref{fig:diff-cores} exhibits the memory utilization in a scale out architecture based on the number of processing units. We notice two things from this figure. We notice in both workloads bandwidth utilization drops as we add more nodes due to additionally adding more memory. We also notice that the graph application itself has a significant affect on the memory utilization as we add more nodes to the system.

As for the intervals of varying compute usage, we sample all of the memory links over a time interval in figure~\ref{fig:BFSWtitter}.
It is key to notice that at any given time some links are at 80\% utilization or lower.
It is also important to note that the flow of the program results in some intervals having much more drastic variance in utilization. 

These lead us to designing a system which can perform a wide array of graph applications, utilize the memory bandwidth to its full potential, decouple memory and compute units, and share memory to achieve maximum memory utilization.

\section{Design of TEGRA}
At a high level, TEGRA is a graph system with a scale up approach made up of many small RISC-V cores adjusted to read and write to hardware message queues in an all-to-all network, with a shared disaggregated memory.
We use general purpose RISC-V cores to allow  workload flexibility over a hardwired accelerator.
On top of general purpose cores, TEGRA takes advantage of three other key components to develop a scalable graph processing system: message passing, memory disaggregation, and heterogeneous memory.

\subsection{Message Passing}
The first key component utilized in TEGRA is message passing. Prior graph frameworks  pass messages from core to core, however this is implemented by expensive loads and stores to main memory. 
This communication consumes extra memory bandwidth which is already limited in graph workloads. This insight leads us to look into alternative methods of core to core communication. Our message passing system is based on Active Messages which uses point to point communication for small messages. 
Active Messages points out that small asynchronous messages targeted to a specific node not only match hardware capabilities better but also allow for the overlap of communication and computation\cite{ActiveMessages}. 
Instead of communicating with other cores through shared memory, a point to point message is sent to a specific destination to tell a core that a vertex needs to be updated. 
These point to point destinations are represented as message queues which are basic FIFO buffers which can be read directly from the core. 
When a core receives a vertex update it triggers the core to access its vertex memory to receive the current property and the address of the edges which need to be used to send new messages. 
Because there is no way to guarantee which message queue a given core may need to access an all to all interconnect is necessary. Each core has its own message queue. 
Message content and sizes depend on the graph workload. In this work the graph workload we are using is Single Source Shortest Path~(SSSP), which finds the shortest path from a single vertex to all the others. In SSSP, messages are 8 bytes containing the vertex ID to be updated and the new weight. 
   
Using this message passing paradigm accomplishes two things: it effectively moves some of the bandwidth burden from the memory bus to the message passing network, and it eliminates copies of data in the cache hierarchy. A traditional system which communicates through memory is going to have copies of updates littered throughout the cache hierarchy as two cores access any given update, one to write and one to read. Point-to-point message passing avoids taking up space in the cache at all from these messages. While message passing helps increase available memory bandwidth it does not enable the decoupling of compute and memory units in the way our second key component, memory disaggregation does.

\subsection{Memory Disaggregation}
\begin{figure}
    \includegraphics[scale=.35]{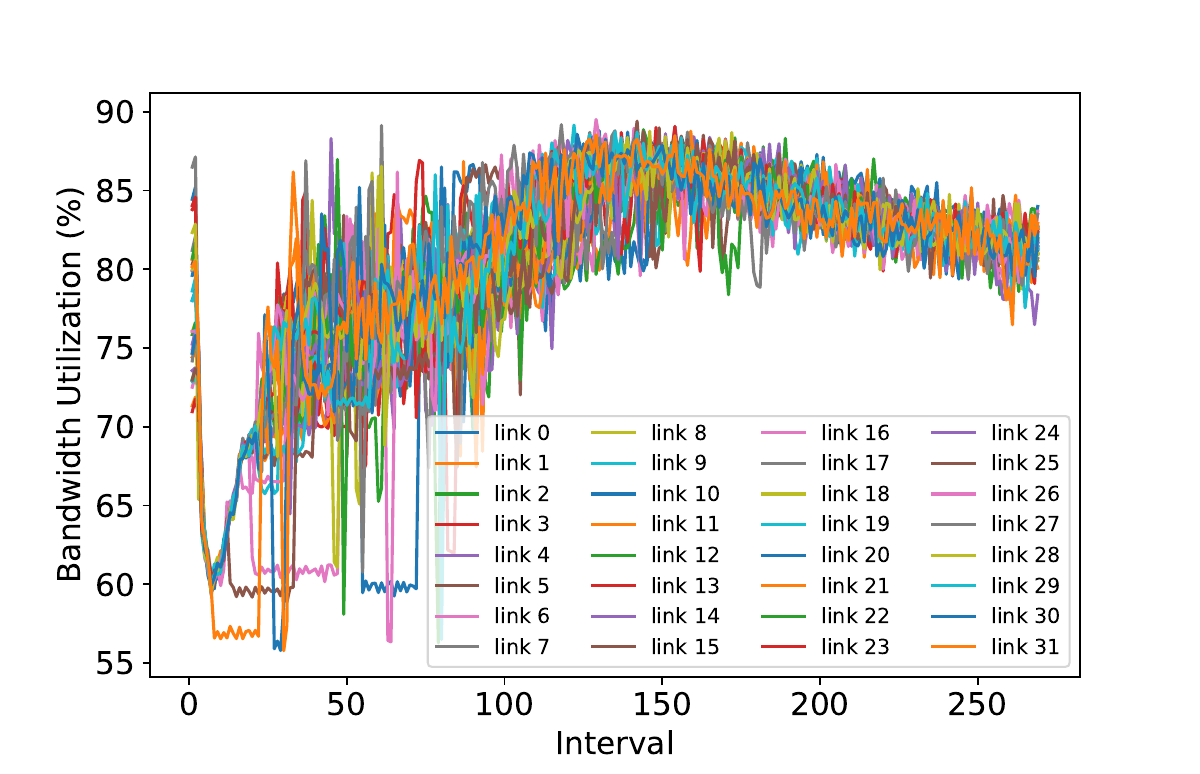}
    \caption{Bandwidth per link over time interval in a scale out graph accelerator. Y-axis starts at 55\%}
    \label{fig:BFSWtitter}
\end{figure}

Recent evolutions in low latency remote memory protocols such as Compute Express Link (CXL)~\cite{CXL} allow for memory to be located farther away from the compute with latency costs around 70-200~ns~\cite{pond}. By locating memory farther away, we are able to connect a system to a larger pool of memory.
TEGRA takes advantage of memory disaggregation as it not only improves the memory capacity of a system but also decouples compute resources from memory resources to allow compute elements to scale independent of memory. By scaling compute units up separate from memory, we can make better use of existing memory bandwidth. In scale-out graph accelerators it is common to see a variation of utilization between any memory channels at a given time interval as seen in figure~\ref{fig:BFSWtitter}.
This indicates that not all cores are seeing the same amount of updates at any given time, as updates drive memory accesses. By using a globally shared memory we are able to share the bandwidth amongst memory channels allowing better bandwidth utilization  regardless of which cores are currently being throttled. While disaggregated memory and compute units allow for potential better bandwidth utilization, the added latency is an important challenge, we address this by utilizing heterogeneous memory.  

\subsection{Heterogeneous Memory}
The third key component of TEGRA is  heterogeneous memory. Vertices and edges exhibit different access patterns and have different requirements from memory. Graphs are commonly represented in a compressed sparse row format. In this format, all the edges of a graph are stored in a sequential array. Each vertex contains an offset as to where in this array it's edges start as well as how many edges it has. Because of this, each time a vertex is accessed, it will access all of it's edges sequentially leading to a small amount of spatial locality. Given that each edge processed requires at least one vertex to be accessed, vertices require higher memory bandwidth than edges. Prior works~\cite{CAL} have shown that HBM performs better for accessing vertices due to their lack of temporal and spatial locality. 
Edges have more (but still limited) spatial locality and take up more memory capacity than vertices which makes disaggregated memory a more suitable option. Because of this the edges are stored in disaggregated DDRx memory.
Now that we have established the design components of TEGRA we must address how the system is programmed.

\subsection{Programming Model}
In this work we will focus on how SSSP can be implemented on TEGRA. We break this down into two algorithms, the message consumer and the message generator. 
The message consumer reads updates from the message queue and accesses vertex memory to see if the update is valid. An update is considered valid if the distance read from the update is shorter than the accessed vertex's distance. 
Upon reading a valid update, the message consumer will update the active list which alerts the message generator which vertex is active. 
The message generator reads from the active list, reading a pointer to where in the edge list to find this vertex's edges, how many edges it has, and the current weight of the vertex. 
Using this information, the message generator begins reading from the edge memory, generating a message for each edge then sending it into the network. 
The message generator and message consumer must run independently to avoid deadlock. For this reason these are each run as separate threads on each core. 
Because these cores communicate to each other with the active list, we need to ensure that either the consumer or generator will not deadlock due to waiting on available resources. This can be done by implementing active list in memory directly or overflowing from a hardware based queue into main memory. 
This guarantees that the message consumers can always continue to read from the message queue preventing a deadlock.

   









\begin{figure}
    \includegraphics[scale=.35]{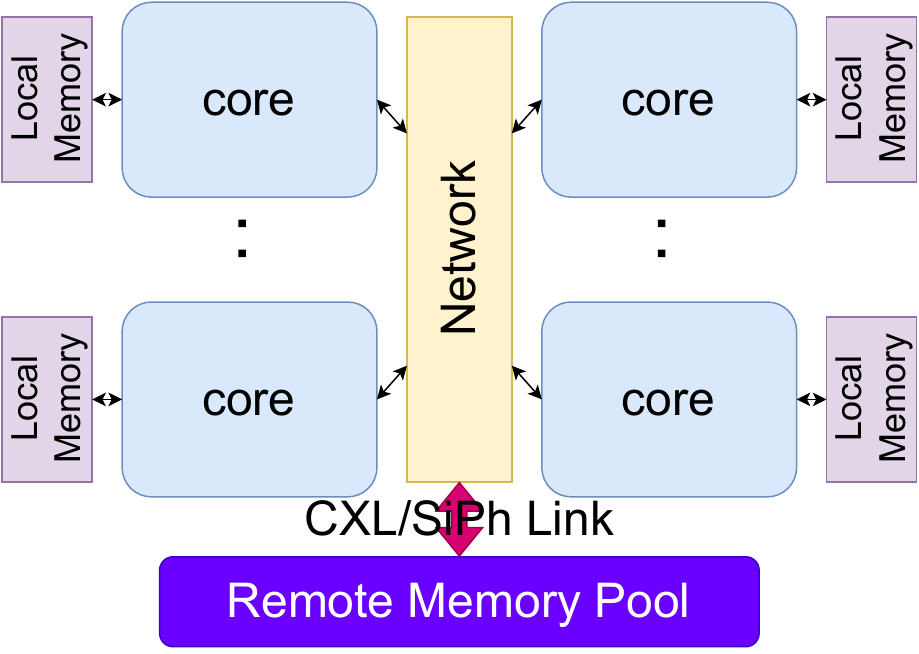}
    \caption{TEGRA Design}
    \label{fig:Tegra}
\end{figure}
\section{Evaluation Methodology}

\begin{figure}
    
    \includegraphics[scale=.35]{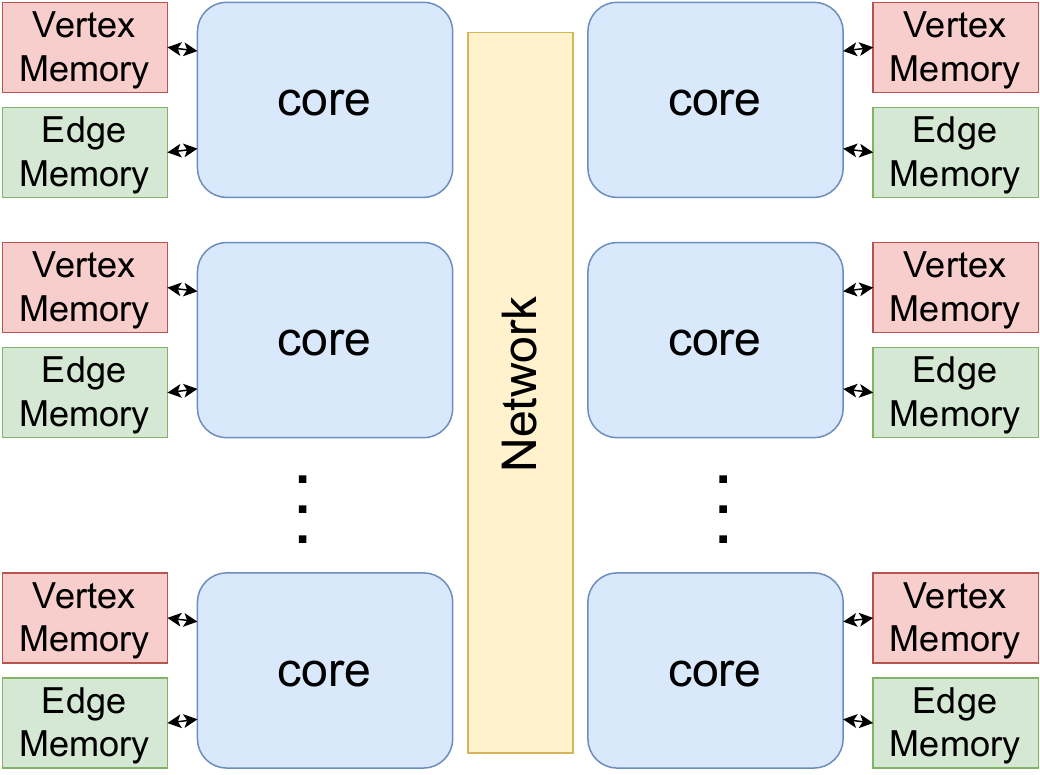}
    \caption{Accelerator System}
    \label{fig:accelerator}
\end{figure}

\begin{figure}
    
    \includegraphics[scale=.35]{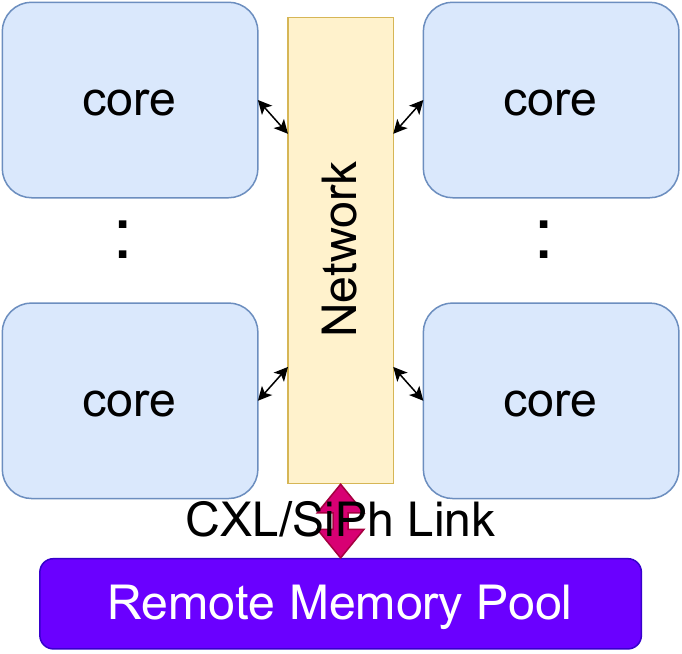}
    \caption{All Disaggregated System}
    \label{fig:Disaggregated}
\end{figure}

To evaluate this system we implement it in the gem5 simulator~\cite{lowepower2020gem5}, using syscall emulation~(SE) mode. We enabled hardware message passing by adding message queues to the system, each core has its own message queue. Upon reading and writing to the message queues' dedicated addresses, we catch the request before it enters the cache hierarchy and forward it to the message queues. We are able to do this by manually mapping the virtual address~(VA) used in our C++ code to the physical address~(PA) ranges of the message queues, which is a gem5 feature not available in full system mode.  We implement SSSP with message passing and run it on three systems to get comparative performance. All three systems utilize the message passing described above and is evaluating the effectiveness of memory placement. The first system is a system using message passing and one shared disaggregated memory seen in Figure
~\ref{fig:Disaggregated}, the second system replicates scale out accelerators where each processing element has its own edge memory and its own vertex memory seen in Figure~\ref{fig:accelerator}, the third system represents TEGRA as seen in Figure \ref{fig:Tegra}. In order to model disaggregated memory, we add a latency of 150~ns to any memory that is disaggregated. In this work we use runtimes of all three systems on the same graph to compare performance.  

For the accelerator like system, we assume two memory channels per core. One DDR4 for edge memory and one HBM2 stack for vertex memory. For our proposed system we assume one stack of HBM2 for vertex memory per core. 

\section{Results}

\begin{figure}
    \includegraphics[scale=.35]{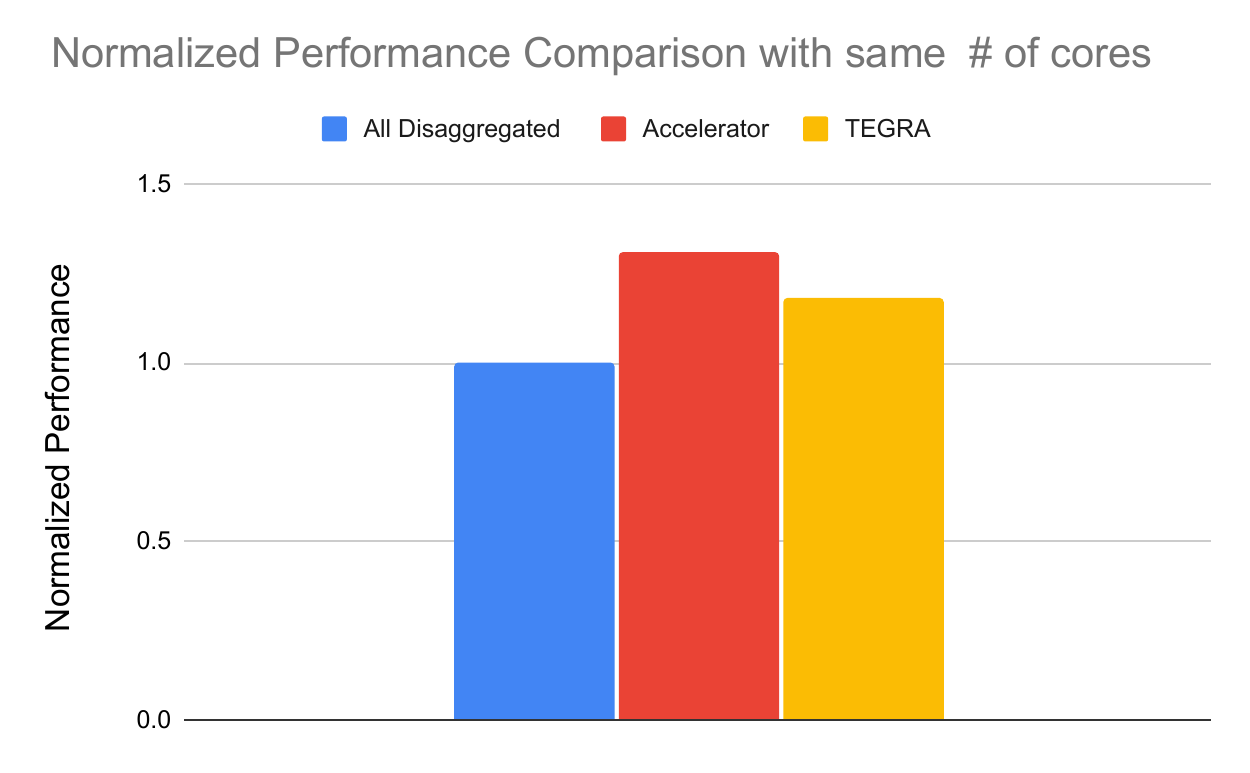}
    \caption{Normalized Performance}
    \label{fig:perf}
\end{figure}

\begin{figure}
    \includegraphics[scale=.35]{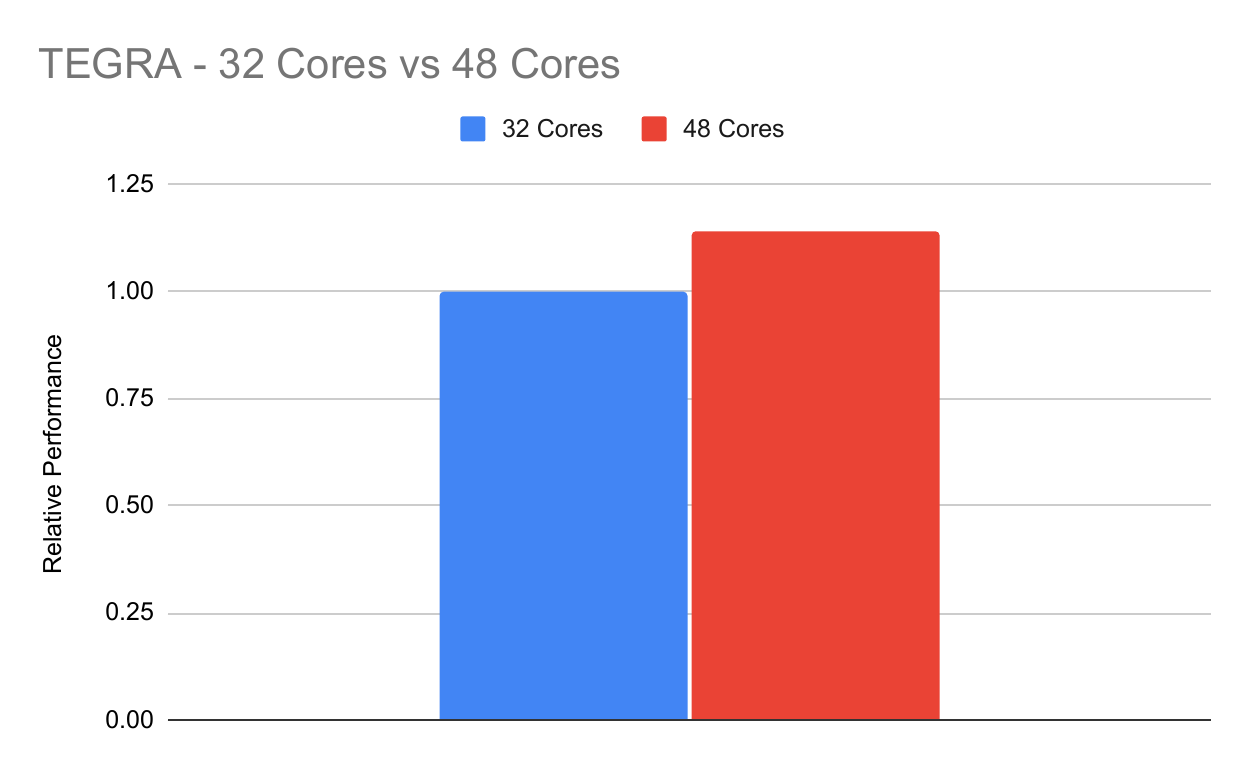}
    \caption{32 cores vs 48 cores in TEGRA}
    \label{fig:32v48}
\end{figure}

Our preliminary results show that in comparison to an all disaggregated system with message passing, the accelerator like system improves performance by 30\% while TEGRA improves performance by 18\% as can be seen in Figure \ref{fig:perf}. We expect the accelerator setup to perform best as it does not suffer the latency of memory disaggregation. 
Given the similarities of systems as far as their message passing systems and binaries go, we can say performance discrepancy comes from the latency of disaggregated memory. This means that keeping vertex memory near provides that 18\% speedup in TEGRA over the all disaggregated system. 
While the accelerator setup outperforms TEGRA and the all disaggregated setup 
both of the other systems allow increasing the number of cores without increasing the amount of edge memory. 
Figure~\ref{fig:32v48} shows TEGRA first with 32 cores, then with 48 cores with the same memory subsystem setup. 
This gives a 13\% improvement to performance showing us more compute units can make use of unutilized memory bandwidth. We expect this performance to improve as we add more cores, however as be begin to reach closer and closer to the peak BW we expect performances to plateau.  

\section{Future Work}
Going forward from here we would like to work on enabling faster simulation of this system in gem5 to enable 1024+ cores and larger graphs for TEGRA. Another key work in the future is adjusting the RISC-V ISA to allow message passing implicitly which would allow this system to be simulated in full system mode within gem5.  

\bibliographystyle{plain}
\bibliography{samples/sample-base}

\doclicenseThis

\end{document}